\documentclass[11pt,twoside]{article}
\usepackage{./asp2014}

\aspSuppressVolSlug
\resetcounters

\begin{document}

\title{The Variable Line Width of Achernar} \author{Th.~Rivinius,$^{1}$
  R.H.D.~Townsend,$^{2}$ D.~Baade,$^{1}$ A.C.~Carciofi,$^{3}$ N.~Leister,$^{3}$
  and S.~\v{S}tefl$^{4\,\dagger}$
\affil{$^{1}$ESO -- European Organisation for Astronomical Research in the
  Southern Hemisphere, Chile \& Germany; email: {\tt triviniu@eso.org}}
\affil{ $^{2}$Department of Astronomy, Univ.\ of Wisconsin-Madison, USA}
\affil{$^{3}$Instituto de Astronomia, Geof\'isica e Ci\^encias Atmosf\'ericas, Universidade
de S\~ao Paulo, Brazil}
\affil{$^{4}$ESO/ALMA -- The Atacama Large Millimeter/Submillimeter Array,
  Chile}
\affil{$^{\dagger}$deceased}}

\paperauthor{Th.~Rivinius}{triviniu@eso.org}{}{ESO - European Organisation for Astronomical Research in the Southern Hemisphere}{}{Vitacura}{}{}{Chile}
\paperauthor{D.~Baade}{dbaade@eso.org}{}{ESO - European Organisation for Astronomical Research in the Southern Hemisphere}{}{Garching}{}{}{Germany}
\paperauthor{S.~\v{S}tefl}{}{}{ESO/ALMA - The Atacama Large Millimeter/Submillimeter Array}{}{Santiago}{}{}{Chile}
\paperauthor{R.H.D.~Townsend}{}{}{Department of Astronomy, Univ.\ of Wisconsin-Madison}{}{Madison/WI}{}{}{USA}
\paperauthor{A.C.~Carciofi}{}{}{Instituto de Astronomia, Geof\'isica e Ci\^encias Atmosf\'ericas}{}{Sao Paulo}{}{}{Brazil}

\begin{abstract}
Spectroscopic observations of Achernar over the past decades, have shown the
photospheric line width, as measured by the rotational parameter $v \sin i$,
to vary in correlation with the emission activity. Here we present new
observations, covering the most recent activity phase, and further archival
data collected from the archives. The $v \sin i$ variation is confirmed.  On
the basis of the available data it cannot be decided with certainty whether
the increased line width precedes the emission activity, i.e. is a signature
of the ejection mechanism, or postdates is, which would make it a signature of
re-accretion of some of the disk-material. However, the observed evidence
leans towards the re-accretion hypothesis.  Two further stars showing the
effect of variable line width in correlation with emission activity, namely
66\,Oph and $\pi$\,Aqr, are presented as well.
\end{abstract}

\section{Introduction}
Recently, the bright southern Be star Achernar ($\alpha$\,Eri, HD\,10144,
B4\,IV) has been found to show variations in the width of its photospheric
lines, as measured by the rotational parameter $v \sin i$
\citep{2013A&A...559L...4R}. The amplitude of this variation is about 15\% of
the value of $v\sin i$ of 250\,km\,s$^{-1}$. This was interpreted by
\citeauthor{2013A&A...559L...4R} with the help of preliminary modeling in the
sense that at least the photospheric layers of Achernar (more specific, at
least the equatorial layers, which contribute most strongly to the line shape
and width) can change their rotational velocity on a relatively short time
scale.

This variation is correlated with the circumstellar emission activity in the
sense that during such activity the line is wider. However, the precise sense
of this correlation could not be determined. The first possiblity is that the
speed-up of the photosphere precedes the activity, and this way could be
understood as a trigger and possibly physical origin of the outburst.  Also
possible, however, is that the speed-up of the photosphere develops only when
parts of the newly formed disk re-accrete onto the star (which is an
unavoidable feature of a disk controlled by viscosity), and in this way is not
cause, but consequence of the circumstellar activity.

New observational data, as well as more data from observing runs predating the
public archive era, were collected to shed light on this question. Available
data for other stars were scanned for similar signatures as seen in Achernar
as well.

\section{Achernar}
\subsection{Archival Data and New Observations}
\begin{figure}[t]
\begin{center}
\includegraphics[width=\textwidth]{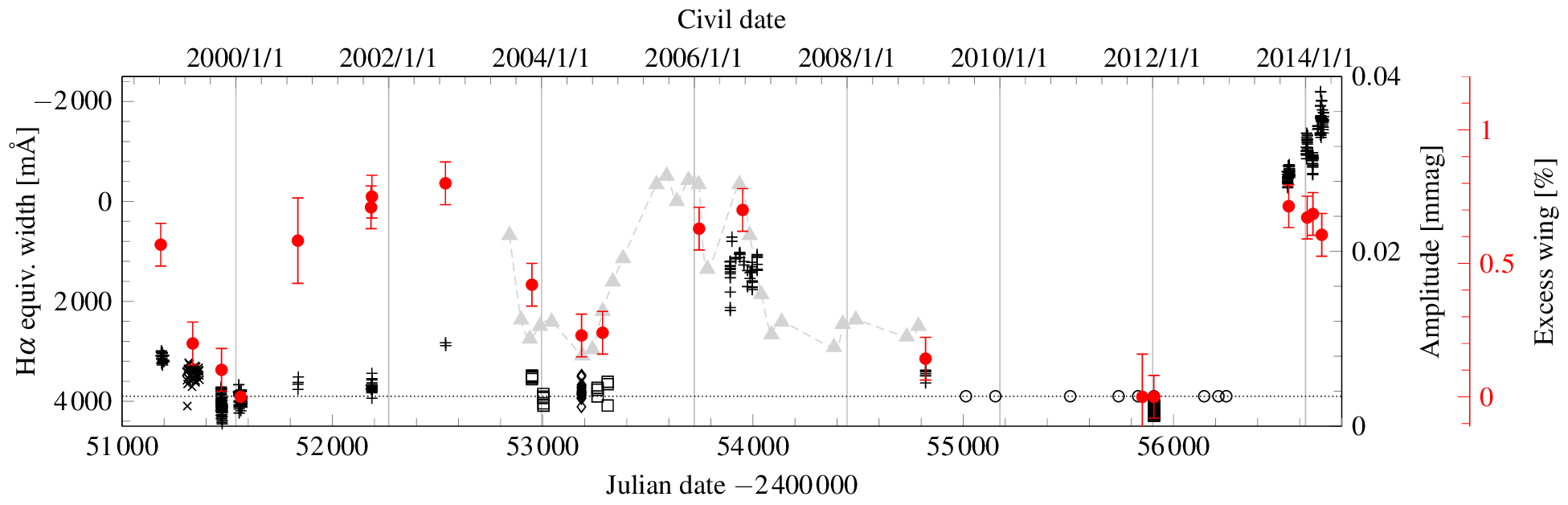}%

\caption{An updated version of Fig.~1 of \citet[][see there for a detailed
    description and references]{2013A&A...559L...4R}, including new data
  mentioned in the text. Filled red disks indicate the strength of the $v sin
  i$ change vs.\ the quiescent (slowly rotating) state, as measured by excess
  wing indentation (Such indentations are present in the blue and red wings of
  residual \ion{He}{i} line profiles, see e.g.\ in the lower panel of
  Fig.~\ref{Rivinius_Achernar_66Oph}, and Fig.~2 of
  \citeauthor{2013A&A...559L...4R} for details of definition and
  measurement), while open symbols and crosses indicate emission activity,
  measured by different instruments. The closed grey triangles mark the
  photospheric pulsation amplitude.}

\label{Rivinius_Achernar_1}
\end{center}
\end{figure}

Additionally to the spectra already shown by \citet{2013A&A...559L...4R},
further data could be acquired. Prior to 2002, FEROS spectra were not
routinely injected into the ESO archive. The star was observed, however,
during the Brazilian time, and these data could be recovered and included in
the dataset \citep{2006A&A...446..643V}. Additional data for the late 2000's
and early 2010's were taken from \citet[][see their Table~3 and
  Fig.~4]{2014A&A...569A..10D}. As mentioned by \citet{2013A&A...559L...4R}, a
new circumstellar activity episode started in January 2013, and Achernar was
observed as a target of opportunity with FEROS mounted at the 2.2m-MPG
telescope at La Silla. A detailed description of the newly acquired data,
archival and recent observations, will be published elsewhere together with a
more complete modeling of the variations.

This additional data were subjected to the same analysis as the previous ones,
outlined in \citet{2013A&A...559L...4R}. In particular, we refer to Fig.~2 of
that work for a description of the residual spectra and the signatures seen in
them. The resulting parameters, excess wing indentation and equivalent width,
are plotted jointly with the previous data in
Fig.~\ref{Rivinius_Achernar_1}.

\subsection{Correlations Between Activty and Rotation}

The most important question is whether the amended data set allows a clearer
view on the causal link behind the observed correlation between emission
activity and the stellar surface rotation. The data now shows two full
emission cycles (2000 to 2004 and 2004 to 2010, cycles 2 and 3), the end
of the previous (cycle 1), and the begining of the most recent ones (cycle 4).

\begin{description}
\item[Cycle 1:] The disk emission dissipates fully. The rotational width also
  goes back to the base value during quiescence, although it seems it reaches
  the base level only slightly after the emission has disappeared.
\item[Cycle 2:] Around MJD\,51\,900 the star does not show any emission, but
  enhancend $v \sin i$. The latest at MJD\,52\,200 a disk starts to
  develop. The disk remains weak and then decays until MJD\,53\,400. However, 
  the rotational width does not fully decay back to its quiescence level.
\item[Cycle 3:] Instead, almost immediately after MJD\,53\,400 a new activity
  cycle starts (seen best in the photometric amplitude data by
  \citet{2011MNRAS.411..162G}. The disk evolves to a stronger emission than in
  cycle 2, and then decays back to zero by about MJD\,55\,200. At this time,
  the rotational enhancement is still marginally present. Note the pulsation
  amplitude has returned to its base level already around MJD\,54\,000, which
  may indicate that dosk feeding has as well ended already then.
\item[Cycle 4:] Achernar remained inactive for several years until
  MJD\,56\,600. During this time, also no enhancement of $v \sin i$ was
  observed. Since the beginning of disk formation after MJD\,56\,600 both
  emission and rotational enhancement are strong.
\end{description}
The above produces a puzzling picture: Cycles 1, 3, and 4 support one picture
in which the rotational enhancement is fully in parallel with the presence of
a disk, which, as will be discussed below, points to the enhancement as a
consequence of the repsence of a disk. Cycle 2, however, supports view that
the rotational enhancement precedes the presence of the disk, which in the
most simple picture would link it to the cause of disk formation, rather that
to disk dissipation. Before discussing this possible contradiction below,
further possible cases of rotational variability will be presented in the next
section.

\section{Other Stars}
\subsection{Disk Decay of 66\,Oph}
\begin{figure}[t]
\begin{center}
\includegraphics[width=\textwidth]{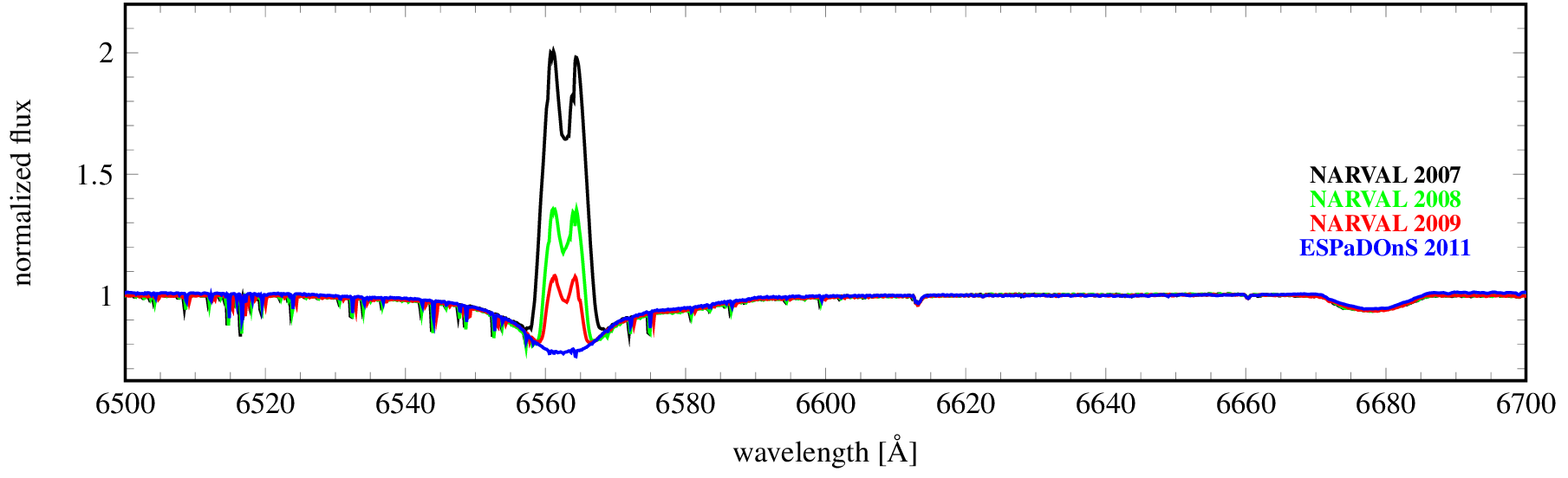}%

\includegraphics[width=\textwidth]{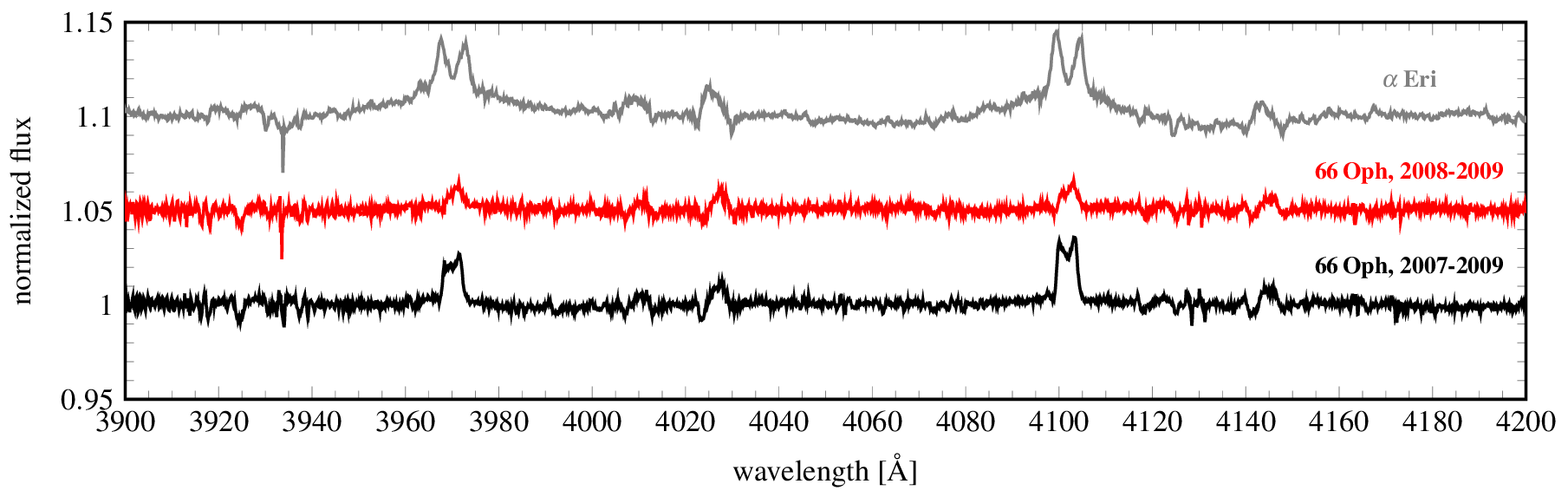}%

\caption{Top: The decay of the H$\alpha$ emission of 66\,Oph, observed from
  2007 (largest emission, but already far away from its maximum several years
  before) to 2011 (no emission). Bottom: Residual spectra of 66\,Oph 2007 and
  2008 vs.\ the low emission state of 2009. A typical Achernar residual
  spectrum for active vs.\ inactive state is shown above for comparison. For a
  more in-depth discussion of the construction of such residual spectra and
  their interpretation see \citet{2013A&A...559L...4R}. }

\label{Rivinius_Achernar_66Oph}
\end{center}
\end{figure}

Between about 1993 and 2011 the H$\alpha$ emission of the Be star 66\,Oph
continuously decayed to the level of zero emission \citep{66Oph_Anatoly}. For
the final few years of this phase archival high quality spectra are available,
which were originally taken for the MiMeS program on Magnetism in Massive
Stars (see Wade et al., this volume) with NARVAL at the TBL (2007, 2008, and
2009) and ESPaDOnS at the CFHT (2011). For each observation multiple frames
were taken in short succession for polarimetric purposes, for this work all
spectra taken during these short epochs were co-added. The upper panel of
Fig.~\ref{Rivinius_Achernar_66Oph} shows this decay. Although brief phases of
injection of fresh material into the circumstellar environement are seen in
these data, the general downward trend is hardly disturbed.

Analoguous to the case of Achernar, residuals were constructed, in which the
state with low emission was taken as a reference.  Unfortunately, the
continuum normalization is not fully consistent between the NARVAL and the
ESPaDOnS data. Therefore, the construction of residuals is constrained to the
NARVAL data to eliminate the effect, and the 2009 NARVAL spectrum is taken as
reference.

The lower panel of Fig.~\ref{Rivinius_Achernar_66Oph} shows the residual
spectra constructed this way, together with a typical residual spectrum of
Achernar. The signatures are very similar, for both the decrase of
circumstellar emission seen in the Balmer lines as well as the change of line
width signature in the \ion{He}{i} and \ion{Ca}{ii} lines. In fact, the most
obvious difference between Achernar and 66 Oph is the lack of broad emission
wings in the Balmer lines and the shell signature in \ion{Ca}{ii}\,3933. The
latter is simply a consequence of 66\,Oph being seen at a more polar
inclination angle, the former possibly due to the fact that the disk in
66\,Oph has been in quasi-steady decay since many years, while in Achernar the
disk density profile, and hence the scattering close to the star, is more
dynamic.

\subsection{Disk Formation of $\pi$\,Aqr}
\begin{figure}[t!]
\begin{center}
\includegraphics[width=\textwidth]{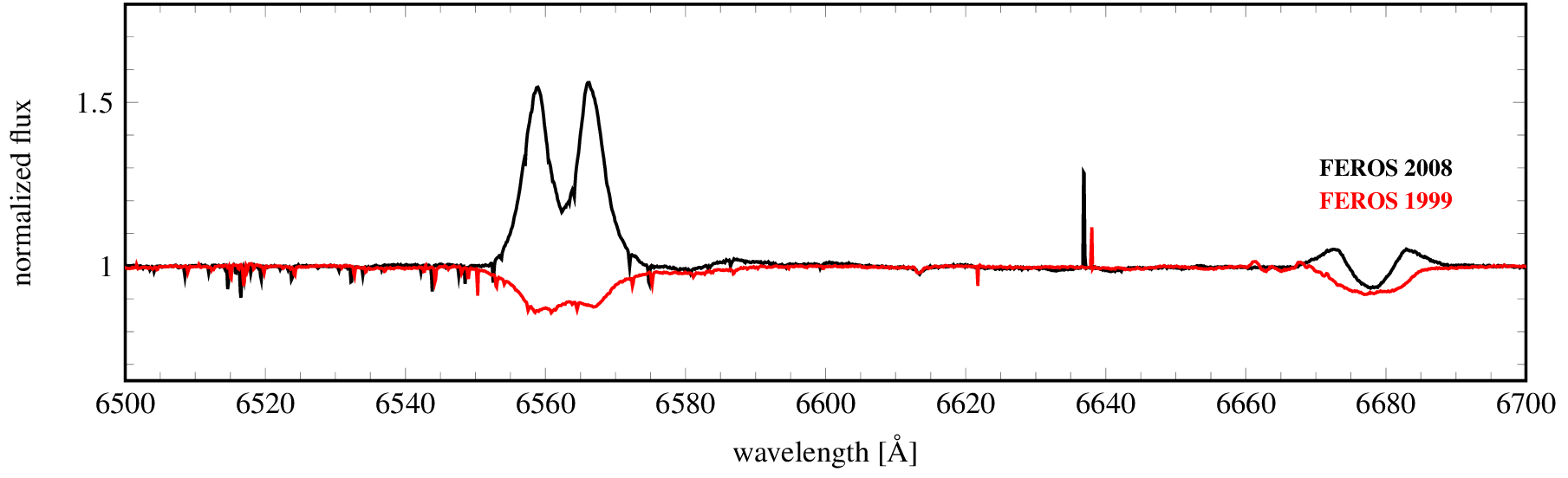}%

\includegraphics[width=\textwidth]{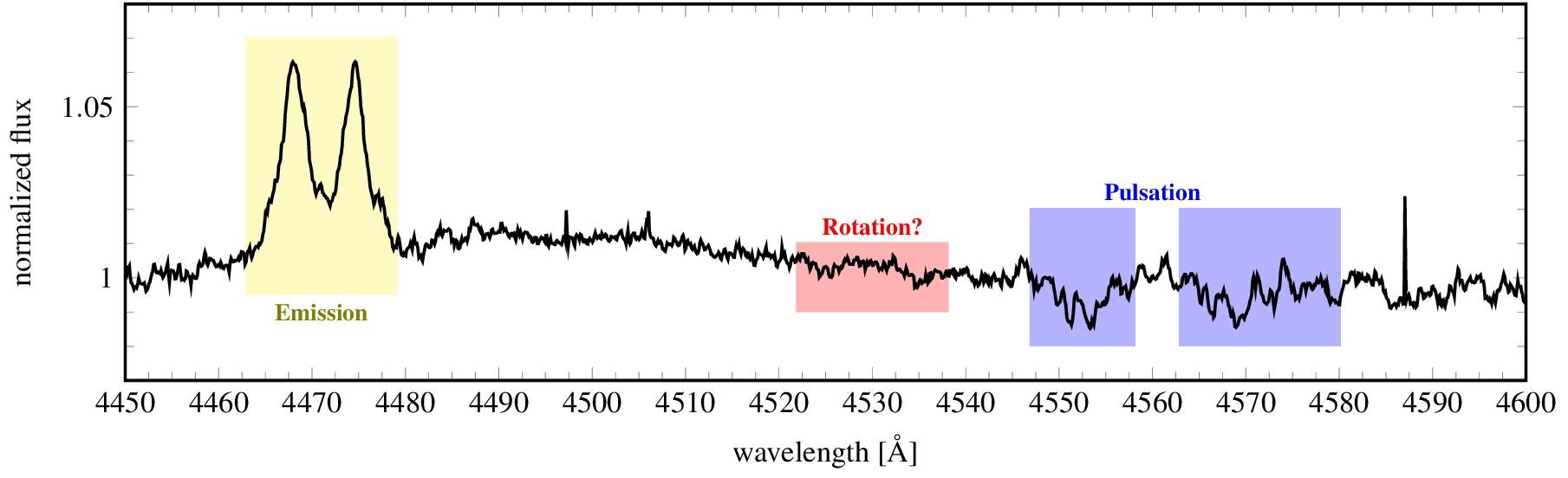}%

\includegraphics[width=\textwidth]{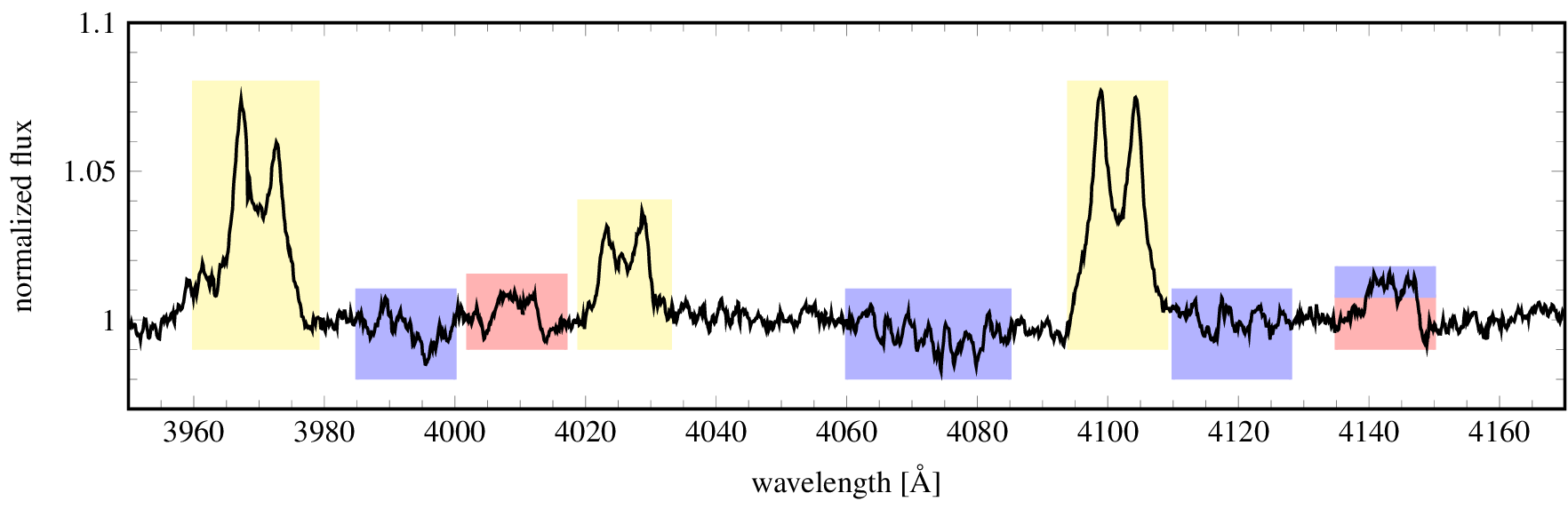}%

\caption{Top: The increase of the H$\alpha$ emission of $\pi$\,Aqr in 2008
  vs.\ 1999. Middle:  Residual spectra region of \ion{He}{i}\,4471,
  \ion{He}{ii}\,4542, and the \ion{Si}{iii} triplet at 4553/68/75, each
  having  a distinct residual signature. 
 Bottom: Residual spectra of the blue wavelength region similar as shown in
 Fig.~\ref{Rivinius_Achernar_66Oph}, with blending of signatures from
 various effects in individual lines.}

\label{Rivinius_Achernar_piAqr}
\end{center}
\end{figure}
$\pi$\,Aqr was observed in a phase of disk build-up.  The disk decayed to
almost emission-less state in the mid 1990's, but did never actually reach
zero emission. Rather, emission remained at a minimum for several years and
$\pi$\,Aqr started to re-build a more massive disk in the mid 2000's
\citep{2010ApJ...709.1306W}. Two high-quality spectra, covering the low state
in 1999 and the active state in 2008, are available in the public archives
(Fig.~\ref{Rivinius_Achernar_piAqr}, upper panel).  Both were taken with
FEROS, which 1999 was mounted at the 1.52m-ESO and 2008 at the 2.2m-MPG
telescopes at La Silla.

Other than 66\,Oph and Achernar, which are mid-type Be stars, $\pi$\,Aqr is an
early type Be star. The photospheric spectrum exhibits lines of \ion{He}{ii},
and when in active state, the spectrum shows \ion{He}{i} lines in emission. It
is further a non-radial $\beta$\,Cephei type pulsator
\citep{2005ASPC..337..294P}. These properties seriously complicate the
detection of subtle signatures in the residual spectra, in particular since
the signatures are most obvious in \ion{He}{i} lines.

The middle panel of Fig.~\ref{Rivinius_Achernar_piAqr} attempts to disentangle
the different dignatures. While the residual profile of \ion{He}{i}\,4471 is
clearly dominated by the change of the emission activity, the residuals of the
\ion{Si}{iii} triplet are hallmarking the non-radial pulsation signature. The
residual signature of \ion{He}{ii}\,4542, however, shows something
different. It has no emission contribution, and as well the variability with
the pulsation period, seen in dynamical spectra similar to the ones shown in
Fig.~2 of \citet{2005ASPC..337..294P}, is very weak and restricted to the line
center. Yet, the outer wings of the line show the similar ``dents'' as the
ones indicating the $v\sin i$ variations in Achernar.

The lower panel illustrates how these three signatures can get blended in
spectral lines. The two Balmer lines have emission contribution only, while
the metallic lines show only pulsational residuals. The \ion{He}{i} lines at
4009, 4021, and 4144 show signatures blended. In \ion{He}{i}\,4009 no emission
signature is seen, but a small pseudo-absorption close to the line center is
due to the pulsation. The overall residual signature of this line, however, is
dominated by the supposed $v\sin i$ variation signature. In \ion{He}{i}\,4144
$v\sin i$ variation and pulsation signature are about equally strong.
The residual signature of \ion{He}{i}\,4021 is dominated by the line emission,
but both the pulsation signature close to the line center as well as the outer
wing indentations due to rotational variability can be detected on a close
look.

Although the case for $\pi$\,Aqr is certainly weaker than that for Achernar or
66\,Oph, the star should still be considered a strong candidate for rotational
variability connected to the disk activity state.

\section{Discussion and Conclusion}

The interpretation of the observed correlations between disk activity and
rotational enhancement is not straightforward. Among the most simple
possibilities is that the rotational enhacenment is due to an acceleration of
equatorial surface material by disk material re-accreting onto the star with
slightly sub-Keplerian velocity. The enhanced velocity of the surface layers
then dissipates by transfering the excess angular momentum back into the star
and maybe as well sidewards into the non-equatorial surface layers as
well. Observations in favour of this scenario are the presence of rotational
enhancement up to the late phases of disk emission, i.e.\ at times by when the
disk feeding should have stopped already long ago (cycle 1 and cycle 3 of
Achernar, 66\,Oph). We note that such a re-accretion inevitably has to occur
for a disk that is governed by viscosity, because there is no other way to
transfer angular momentum outwards than to take it away from particles that
subsequently fall back onto the star. This process starts as soon as the disk
exists and only ends when the disk is fully dissipated. The presence of
rotational enhancement without a disk, such as in cycle 2, is an argument
against this simple process, unless one speculates about a brief (and not
observed) emission epsiode just before the observations. 

A second simple possibility is that angular momentum is transported towrds the
surface by processes from inside the star, e.g.\ the pulsation, until the
surface nears the critical limit, at which point this angular momentum is shed
by the formation of a disk. Then the rotational enhancement should be out of
phase with the disk presence, by about 1/4 of a cycle. While cycle 2 does
point to such a phase difference, the absence of any rotational enhancement in
the quiescent period before cycle 4 speaks against it. Just as in the above
hypothesis cycle 2 could be speculated to be due to a brief activity period,
the same could be speculated for the rotational enahncements seen in the late
phases of a cycle, namely to be due to brief, but otherwise undetected,
activity phases.

Summing up the evidence for the above two possibilities, the first seemes to
require fewer additional assumptions (just one unseen brief activity phase),
while the second not only demands several of those in several stars, but still
leaves the quiescence before the activity onset in cycle 4 unexplained. 

Adding an additional trigger process, that keeps the rotational enhancement
due to pulsation at bay during the inactive phases, and speeds up the surface
only just before the outburst does only move the problem one layer down: It is
exactly the appeal of the angular-momentum-spillover hypothesis to remove the
need for such a trigger. Although not entirely conclusive, the sum of evidence
seems to weigh towards the re-accretion hypothesis.

In any case, the discovery of the same type of signature in 66\,Oph, and
likely as well in $\pi$\,Aqr, makes it worthwhile testing the hypothesis that
the rotational variability as seen in Achernar as a property of all Be stars.

\acknowledgements This work made use of observations collected at the European
Southern Observatory at La Silla and Paranal, Chile, Prog. IDs: 62.H-0319,
64.H- 0548, 072.C-0513, 073.C-0784, 074.C-0012, 073.D-0547, 076.C-0431,
077.D-0390, 077.D-0605, and the technical program IDs 60.A-9120 and 60.A-9036,
of observations obtained at the Canada-France-Hawaii Telescope (CFHT) which is
operated by the National Research Council of Canada, the Institut National des
Sciences de l'Univers (INSU) of the Centre National de la Recherche
Scientifique (CNRS) of France, and the University of Hawaii, of observations
obtained using the Narval spectropolarimeter at the Observatoire du Pic du
Midi (France), which is operated by CNRS/INSU and the University of Toulouse,
and of observations acquired with FEROS at the 2.2m-MPG telescope located on
La Silla, Prog. IDs: 089.A-9032(A) and 091.A-9032(D).

\bibliography{alp}  

\end{document}